\documentstyle[12pt]{article}
\begin{document}
\baselineskip=20pt

\pagenumbering{arabic}

\null\vspace{2.0cm}

\begin{center}
{\Large\sf Renormalization of the Electroweak Theory
in the Background Field Method}
\\[10pt]
\vspace{2.ex}
\vspace{1.5cm}

{$\underline{\rm Xiaoyuan~Li{}^a}$ and Yi Liao${}^{\rm b}$}

\vspace{3.5ex}
{a \it Institute of Theoretical Physics, Chinese Academy of Sciences\\
P.O.Box 2735, Beijing 100080, P.R.China \\}
{b \it Department of Modern Applied Physics, Tsinghua University\\
Beijing 100084, P.R.China}
\vspace{5.0ex}

             {\bf Abstract}
\end{center}

The applicability of the background field method
in spontaneously broken gauge theories is examined with
new features emphasized. An explicit one loop analysis
in the electroweak theory shows that the method can be consistently
implemented in the on-shell renormalization scheme, and that the choice of
the background gauge cannot be arbitrary and must be fixed
in the Landau gauge if one calculates scattering amplitudes involving
unphysical Goldstone bosons. Some possible applications
are also briefly indicated.

\begin{flushleft}
\end{flushleft}
\baselineskip=24pt

\newpage
Although the S matrix keeps gauge invariant in gauge theories,
this property is lost at intermediate stages
in the conventional procedure of quantization.
The background field method ( BFM ) is a technique which can
preserve explicit gauge invariance at intermediate stages,
so that Green functions satisfy the naive Ward identities
which are more restrictive than the Slavnov-Taylor identities
from BRS invariance and therefore lead to a simpler renormalization
structure.

The method was developed some time ago for
unbroken gauge theories${}^{\cite{Abbotta}}$, in a manner
analogous to the conventional one and The applicability of the background field
method
in spontaneously broken gauge theories is examined with
new features emphasized. An explicit one loop analysis
in the electroweak theory shows that the method can be consistently
implemented in the on-shell renormalization scheme, and that the choice of
the background gauge cannot be arbitrary and must be fixed
in the Landau gauge if one calculates scattering amplitudes involving
unphysical Goldstone bosons. Some possible applications
are also briefly indicated.
\\applicable to all orders in perturbation theory.
However, a parallel work, in the case of spontaneously broken
gauge theories in general and the electroweak theory in particular,
has not been available to our knowledge
\footnote{The method was used to study symmetry restoration
in spontaneously broken gauge theories in a constant electromagnetic
background field in Ref. $\cite{Shore}$, which however does not
apply to the construction of Green functions. A formally similar
construction also appeared in Ref. $\cite{Weinberg}$ in the
context of the gauge coupling running of the QCD and electroweak
theory in grand unified theories. Recently the method was compared
to the pinch technique ${}^{\cite{Pinch}}$ in Ref. $\cite{Denner}$
which however only treats bare quantities.}.
Although the formal construction in the two cases is similar,
some subtle points associated with renormalization may be
easily missed since bare quantities are usually involved in
formal manipulations. An explicit study is therefore necessary
for the resolution of these points. Indeed, spontaneously broken
gauge theories differ from unbroken ones mainly in their
patterns of symmetry realization. As will be pointed out below,
the constraints on renormalization constants imposed by the
BFM are not apparently understandable in spontaneously broken
gauge theories as in unbroken theories. Furthermore, renormalization
constants also depend on subtraction schemes used.
While in QCD mass-independent
subtractions ( e.g. the minimal subtraction and its modifications )
are usually used, in the electroweak theory many physical scales
are involved and mass-dependent subtractions ( e.g. the on-shell
renormalization scheme ) are usually favoured. It is not
clear at all whether such subtractions are consistent with the
constraints imposed by the BFM in spontaneously broken gauge theories.
It is probable that the consistent implementation of the BFM
inversely picks out appropriate subtraction schemes.
Another feature not encountered in the QCD case is that
the gauge choice for background gauge fields cannot be arbitrary
and must be fixed in the Landau gauge if one calculates scattering
amplitudes involving unphysical Goldstone bosons, while in QCD
the gauge choices for quantum and background gauge fields are
independent and can be both arbitrary.

In this letter we study the renormalization of the electroweak theory
in the BFM by an explicit one loop analysis. Certainly, the full
power of the method can be best enjoyed in higher order calculations,
yet the feasibility of the method must begin at the one loop level
and some features can also be glimpsed at this level. We will show
that the on-shell renormalization scheme is consistent with the BFM
and that this consistency can also be set up by examining the
renormalized Ward identities. Finally we summarize our results
and briefly mention some possible applications.

We begin with the formal construction of the bare Lagrangian
${}^{\cite{Abbotta}}$. The generating functional for connected
Green functions of background fields, $W[J,\hat{f}]$, is defined
by
\begin{equation}
\displaystyle
\exp(iW[J,\hat f])=\int (Df~D\omega~D\overline{\omega})
                    \exp~i\int d^4 x[{\cal L}_{\rm classical}(f+\hat f)
                     +{\cal L}_{\rm g.f.}+{\cal L}_{\rm FP}+Jf],
\end{equation}
where $\hat f$ collectively stands for the generic background field
and $f$ the generic quantum field to be integrated over,
$\omega$ and $\overline{\omega}$ are Faddeev-Popov ghosts.
Note that the external source $J$ is introduced only for the $f$ field.
The gauge fixing term ${\cal L}_{\rm g.f.}$ is so chosen as to
preserve the background gauge invariance of $W[0,\hat f]$
while breaking the quantum gauge invariance. The background
field effective action is defined by the Legendre transform,
\begin{equation}
\begin{array}{l}
\displaystyle\Gamma [\tilde f,\hat f]=W[J,\hat f]-\int d^4 x J\tilde f,\\
{}~~~~~~~~~~\displaystyle\tilde f=\frac{\delta W}{\delta J}.
\end{array}
\end{equation}
$\Gamma [0,\hat f]$ is the gauge invariant effective action that
one computes in the BFM. When augmented by a gauge-fixing
term for background gauge fields, it can be used to generate
the S matrix by constructing trees using its vertices
and propagators${}^{\cite{Abbottb}}$.

Now we restrict ourselves to the electroweak theory of
$SU(2)\otimes U(1)$. For simplicity only the bosonic sector
is included. The inclusion of fermions is straightforward and does
not pose any new problems since fermions constitute a gauge
invariant subset by themselves and there is no need to discriminate
quantum from background fields. We write,
\begin{equation}
\begin{array}{l}
\displaystyle{\cal L}_{\rm classical}
             ={\cal L}_{\rm YM}+{\cal L}_{\rm scalars},\\
\displaystyle{\cal L}_{\rm YM}=-\frac{1}{4}W^a_{\mu\nu} W^{a,\mu\nu}
                               -\frac{1}{4}B_{\mu\nu}B^{\mu\nu},\\
\displaystyle ~~~~~~W^a_{\mu\nu}
                   =\partial_{\mu}W^a_{\nu}-\partial_{\nu}W^a_{\mu}
                    +g_2 \epsilon^{abc} W^b_{\mu}W^c_{\nu},\\
\displaystyle ~~~~~~B_{\mu\nu}
                   =\partial_{\mu}B_{\nu}-\partial_{\nu}B_{\mu},\\
\displaystyle {\cal L}_{\rm scalars}
             =(D_{\mu}\Phi)^{\dagger} (D^{\mu}\Phi)
              +\mu^2\Phi^{\dagger}\Phi
              +\lambda (\Phi^{\dagger}\Phi)^2,\\
\displaystyle ~~~~~~D_{\mu}\Phi
                   =(\partial_{\mu}-ig_2 \frac{\tau^a}{2}W^a_{\mu}
                    -ig_1 \frac{1}{2}B_{\mu})\Phi,\\
\displaystyle ~~~~~~\Phi=\left( \begin{array}{c}
                                  \phi^+ \\
                                  \frac{1}{\sqrt{2}}(\phi_1+v+i\phi_2)
                                \end{array}
                         \right)~~,
\end{array}
\end{equation}
where the usual notations for fields and parameters have been used.
${\cal L}_{\rm classical}(f+\hat f)$ is obtained from the above by the
replacements,
\begin{equation}
\begin{array}{l}
\displaystyle W^a_{\mu}\longrightarrow W^a_{\mu}+\hat{W^a_{\mu}},~~~
              B_{\mu}\longrightarrow B_{\mu}+\hat{B_{\mu}},\\
\displaystyle \Phi \longrightarrow \Phi+\hat{\Phi},~~~
              \Phi=\left( \begin{array}{c}
                           \phi^+ \\
                           \frac{1}{\sqrt{2}}(\phi_1+i\phi_2)
                          \end{array}
                   \right),~~~
\displaystyle \hat{\Phi}=\left( \begin{array}{c}
                                 \hat{\phi}^+ \\
                                 \frac{1}{\sqrt{2}}
                                   (\hat{\phi}_1+v+i\hat{\phi}_2)
                                \end{array}
                         \right).
\end{array}
\end{equation}
Note that the background scalar $\hat{\Phi}$ develops a non-zero VEV
while the quantum scalar $\Phi$ does not. The invariance
of $W[0,\hat f]$ ( and thus $\Gamma[0,\hat f]$ ) under gauge
transformations of the $\hat f$ requires that ${\cal L}_{\rm g.f.}$
( and thus ${\cal L}_{\rm FP}$ ) be invariant under the union of gauge
transformations of the $\hat f$ and appropriate changes of the
integration variables $f$. A straightforward generalization of the
so-called `` background field $R$ gauge '' in QCD ${}^{\cite{Abbotta}}$
and the conventional $R_{\xi}$ gauge in the electroweak theory
${}^{\cite{Li}}$ leads to the following construction
${}^{\cite{Shore}\cite{Weinberg}\cite{Einhorn}\cite{Denner}}$,
\begin{equation}
\begin{array}{l}
\displaystyle {\cal L}_{\rm g.f.}
             =-\frac{1}{2\xi_2}[(\partial_{\mu} \delta^{ac}
              +g_2 \epsilon^{abc} \hat{W}^b_{\mu})W^{c,\mu}
              -ig_2 \xi_2 \frac{1}{2}
              (\hat{\Phi}^{\dagger}\tau^a \Phi
              -\Phi^{\dagger}\tau^a \hat{\Phi})]^2 \\
\displaystyle~~~~~~~~-\frac{1}{2\xi_1}[\partial_{\mu}B^{\mu}
                                      -ig_1 \xi_1 \frac{1}{2}
                                       (\hat{\Phi}^{\dagger} \Phi
                                      -\Phi^{\dagger} \hat{\Phi})]^2.
\end{array}
\end{equation}
Note that the two terms in the first pair of square brackets transform
separately in the adjoint representation under the aforementioned
union. ${\cal L}_{\rm FP}$ is then determined by using quantum gauge
transformations that keep ${\cal L}_{\rm classical}(f+\hat{f})$
invariant.

To go beyond tree level, renormalizations have to be carried out.
The only renormalizations required in the BFM are those of physical
parameters, background fields and gauge parameters $\xi_i$
${}^{\cite{Abbotta}}$. However, for the purpose of calculating
$1$PI functions of background fields to one loop, even the
renormalization of $\xi_i$ is not required since $\xi_i$ appears
only in vertices that are at least quadratic in quantum fields.
For the other renormalizations we take as usual${}^{\cite{Hollik}}$,
\begin{equation}
\begin{array}{l}
\displaystyle\hat{B}_{\mu} \longrightarrow (Z^B_2)^{1/2}\hat{B}_{\mu},~~~
\hat{W}^a_{\mu} \longrightarrow (Z^W_2)^{1/2}\hat{W}^a_{\mu},\\
\displaystyle(\hat{\phi}^{\pm},\hat{\phi}_1,\hat{\phi}_2)
      \longrightarrow (Z_{\phi})^{1/2}
(\hat{\phi}^{\pm},\hat{\phi}_1,\hat{\phi}_2),\\
\displaystyle g_1 \longrightarrow Z^B_1(Z^B_2)^{-3/2} g_1,~~~
g_2 \longrightarrow Z^W_1(Z^W_2)^{-3/2} g_2,\\
\displaystyle\mu^2 \longrightarrow (Z_{\phi})^{-1}(\mu^2-\delta \mu^2),~
\lambda \longrightarrow (Z_{\phi})^{-2} Z_{\lambda}\lambda,~
v \longrightarrow (Z_{\phi})^{1/2} (v-\delta v).
\end{array}
\end{equation}
Explicit gauge invariance of $\Gamma[0,\hat{f}]$ requires
that covariant derivatives be renormalized in the following way, e.g. ,
\begin{equation}
\begin{array}{l}
\displaystyle\hat{D}_{\mu}\hat{\Phi}
=(\partial_{\mu}-ig_1 \frac{1}{2}\hat{B}_{\mu}
                -ig_2 \frac{\tau^a}{2}\hat{W}^a_{\mu})\hat{\Phi}
\longrightarrow
(Z_{\phi})^{1/2}\hat{D}_{\mu}\hat{\Phi},
\end{array}
\end{equation}
where, on the rhs, $v$ has been replaced by $v-\delta v$ and
all quantities in $\hat{D}_{\mu}\hat{\Phi}$ are renormalized
or finite. But this is possible only if
\begin{equation}
\displaystyle
Z^W_1=Z^W_2,~~~Z^B_1=Z^B_2,
\end{equation}
\begin{equation}
\delta v={\rm finite}.
\end{equation}
Eqn. $(8)$ is not self-evident, at least for the finite
parts since $Z_2$'s are generally related to wavefunction
renormalizations while $Z_1$'s go into mass renormalizations.
( This would be clearer if $Z_1$ were replaced
by $Z_1(Z_2)^{3/2}$. ) The similar situation does not occur
in QCD where the symmetry is unbroken. Nevertheless, we
will show explicitly that the constraints of Eqns. $(8)$ and $(9)$
are satisfied at least in the on-shell renormalization scheme
so that naive Ward identities are indeed saturated by $1$PI
functions of background fields.

The Feynman rules and counterterms can now be written down.
Due to the nonlinearity of gauge condition functions and the
discrimination between quantum and background fields, the
Feynman rules are different from those in the conventional
approach. It is worth mentioning that the background Goldstone
bosons are massless since the background gauge has not been
fixed. To avoid tree level $Z_{\mu}-A_{\mu}$ mixing we assume
$\xi_1=\xi_2=\xi$. For simplicity we work in the quantum
't Hooft -- Feynman gauge, $\xi=1$. We found that calculations
in the $\xi=1$ gauge in the BFM are much simpler than in the
conventional approach. The renormalization constants are
determined by the following conditions
\footnote{The following conventions are assumed :
(1) $\Sigma$'s or $\Gamma$'s with a hat, a tilde or nothing
are respectively renormalized, counterterm or unrenormalized
one loop contributions;
(2) factor $i$ has been separated out from $\Sigma$'s or $\Gamma$'s;
(3) for gauge bosons $\Sigma_T$ refers to the
coefficient of $g_{\mu\nu}$;
(4) momenta are taken to be incoming. } ${}^{\cite{Hollik}}$:
\begin{enumerate}
  \item vanishing tadpole $$\hat{T}=0$$

  \item on-shell definition of masses $${\rm Re}\hat{\Sigma}^{\hat{W}}_T
(M^2_W)
                             ={\rm Re}\hat{\Sigma}^{\hat{Z}}_T (M^2_Z)
                             ={\rm Re}\hat{\Sigma}^{\hat{\phi}_1}(M^2_H)=0$$

  \item explicit $U(1)_{\rm e.m.}$ symmetry
       $$\displaystyle [\frac{1}{p^2}\hat{\Sigma}^{\hat{A}}_T (p^2)]
                       _{p^2=0}=0,~~~
        \hat{\Sigma}^{\hat{A}\hat{Z}}_T (0)=0$$

  \item unity residue of the $\hat{\phi}^{\pm}$ propagator
       $$\displaystyle [\frac{1}{p^2}\hat{\Sigma}^{\hat{\phi}^{\pm}}(p^2)]
                       _{p^2=0}=0$$

  \item electromagnetic coupling defined in the Thomson limit
        \footnote{Only the charge interaction part is relevant.}
       $$\displaystyle [\hat{\Gamma}^{\hat{A}\hat{\phi}^{+}\hat{\phi}^-}
                        _{\mu}(p,p_+,p_-)]_{p^2,p^2_{\pm} \rightarrow 0}
       =e(p_+-p_-)_{\mu}~.$$
\end{enumerate}

The electromagnetic coupling definition requires unity residues of the
photon and the reference particle propagators. Since there is no extra
freedom to require unity residue for the $\hat{W}^{\pm}$ propagator
when this has been done for the $\hat{A}$ propagator, we have used
$\hat{\phi}^{\pm}$ as the reference particle and required unity residue
of its propagator instead of the $\hat{\phi}_1$ propagator. If
we include fermions we may use, e.g. the electron,
as the reference particle and obtain the same result for the charge
renormalization.

The counterterm contributions to the above quantities are,
\begin{equation}
\begin{array}{l}
\displaystyle\tilde{T}=-v[\delta \mu^2
            +\lambda v^2 (\delta Z_{\lambda}-2\frac{\delta v}{v})],\\
\displaystyle\tilde{\Sigma}^{\hat{W}}_{\mu\nu}
=-\delta Z^W_2(p^2 g_{\mu\nu}-p_{\mu}p_{\nu})
 +g_{\mu\nu}(\delta M^2_W+M^2_W \delta Z^W_2),\\
\displaystyle\tilde{\Sigma}^{\hat{Z}}_{\mu\nu}
=-\delta Z^Z_2(p^2 g_{\mu\nu}-p_{\mu}p_{\nu})
 +g_{\mu\nu}(\delta M^2_Z+M^2_Z \delta Z^Z_2),\\
\displaystyle\tilde{\Sigma}^{\hat{A}}_{\mu\nu}
=-\delta Z^A_2(p^2 g_{\mu\nu}-p_{\mu}p_{\nu}),\\
\displaystyle\tilde{\Sigma}^{\hat{A}\hat{Z}}_{\mu\nu}
=-\delta Z^{AZ}_2(p^2 g_{\mu\nu}-p_{\mu}p_{\nu})
 +g_{\mu\nu}M^2_Z(\delta Z^{AZ}_1-\delta Z^{AZ}_2),\\
\displaystyle\tilde{\Sigma}^{\hat{\phi}_1}
=\delta Z_{\phi}p^2-(\delta M^2_H+M^2_H \delta Z_{\phi}),\\
\displaystyle\tilde{\Sigma}^{\hat{\phi}_2}=\tilde{\Sigma}^{\hat{\phi}^{\pm}}
=\delta Z_{\phi}p^2
-[\delta \mu^2+\lambda v^2 (\delta Z_{\lambda}-2\frac{\delta v}{v})],\\
\displaystyle\tilde{\Gamma}^{\hat{A}\hat{\phi}^{+}\hat{\phi}^{-}}_{\mu}
                    (p,p_+,p_-)
=e(p_+-p_-)_{\mu}[\delta Z_{\phi}+\delta Z^A_1-\delta Z^A_2
  +\frac{c^2-s^2}{2cs}(\delta Z^{AZ}_1-\delta Z^{AZ}_2)].
\end{array}
\end{equation}
We will not list the explicit expressions of the unrenormalized
one loop contributions. In determining renormalization constants
we note the following crucial properties:
\begin{equation}
\begin{array}{l}
\displaystyle\Sigma^{\hat{A}}_{\mu\nu} \propto
                      (p^2 g_{\mu\nu}-p_{\mu}p_{\nu}),\\
\displaystyle\Sigma^{\hat{A}\hat{Z}}_{\mu\nu} \propto
                      (p^2 g_{\mu\nu}-p_{\mu}p_{\nu}),\\
\displaystyle\Sigma^{\hat{\phi}_2}(0)=\Sigma^{\hat{\phi}^{\pm}}(0)
=\delta\mu^2+\lambda v^2(\delta Z_{\lambda}-2\frac{\delta v}{v}),\\
\displaystyle[\Gamma^{\hat{A}\hat{{\phi}^{+}}\hat{{\phi}^-}}_{\mu}
              (p,p_+,p_-)]_{p^2,p^2_{\pm} \rightarrow 0}
             =-e(p_+-p_-)_{\mu} \delta Z_{\phi}.
\end{array}
\end{equation}
It is clear that Eqn. $(8)$ is satisfied and the unphysical Goldstone
bosons $\hat{\phi}_2$ and $\hat{\phi}^{\pm}$ keep massless to one loop.
$\delta v /v$ is determined from $\delta M^2_W,~\delta M^2_Z,~
\delta Z^A_2$ and $\delta Z_{\phi}$ and the result is finite.

The self-consistency of the BFM in the on-shell scheme may also
be checked by examining Ward identities. We derive Ward identities
directly for renormalized $1$PI functions. In this procedure,
Eqns. $(8)$ and $(9)$ are necessary to obtain genuine renormalized
gauge transformations from the bare ones. The gauge invariance
of $\Gamma [0,\hat{f}]\equiv \hat{\Gamma}$ ( Here $\hat{~}$ means
both `` background '' and `` renormalized '' ) gives,
\begin{equation}
\begin{array}{l}
\displaystyle\frac{1}{g_2}\partial_{\mu}
     \frac{\delta \hat{\Gamma}}{\delta \hat{W}^+_{\mu}}
+i(c\hat{Z}_{\mu}+s\hat{A}_{\mu})
     \frac{\delta \hat{\Gamma}}{\delta \hat{W}^+_{\mu}}
-i\hat{W}^-_{\mu}
     (c\frac{\delta \hat{\Gamma}}{\delta \hat{Z}_{\mu}}
      +s\frac{\delta \hat{\Gamma}}{\delta \hat{A}_{\mu}}) \\
\displaystyle-\frac{i}{2}(v-\delta v+\hat{\phi}_1 +i\hat{\phi}_2)
     \frac{\delta \hat{\Gamma}}{\delta \hat{\phi}^+}
+\frac{i}{2}\hat{\phi}^-
     \frac{\delta \hat{\Gamma}}{\delta \hat{\phi}_1}
-\frac{1}{2}\hat{\phi}^-
     \frac{\delta \hat{\Gamma}}{\delta \hat{\phi}_2}=0,\\
\end{array}
\end{equation}
\begin{equation}
\begin{array}{l}
\displaystyle\frac{1}{\sqrt{g^2_1+g^2_2}}\partial_{\mu}
     \frac{\delta \hat{\Gamma}}{\delta \hat{Z}_{\mu}}
+ic^2(\hat{W}^-_{\mu}\frac{\delta\hat{\Gamma}}{\delta\hat{W}^-_{\mu}}
     -\hat{W}^+_{\mu}\frac{\delta\hat{\Gamma}}{\delta\hat{W}^+_{\mu}})\\
\displaystyle +\frac{i}{2}(c^2-s^2)
      (\hat{\phi}^-\frac{\delta\hat{\Gamma}}{\delta\hat{\phi}^-}
      -\hat{\phi}^+\frac{\delta\hat{\Gamma}}{\delta\hat{\phi}^+})
-\frac{1}{2}\hat{\phi}_2 \frac{\delta \hat{\Gamma}}{\delta\hat{\phi}_1}
+\frac{1}{2}(v-\delta v+\hat{\phi}_1)
       \frac{\delta \hat{\Gamma}}{\delta\hat{\phi}_2}=0,\\
\end{array}
\end{equation}
\begin{equation}
\begin{array}{l}
\displaystyle\frac{1}{e}\partial_{\mu}
       \frac{\delta \hat{\Gamma}}{\delta \hat{A}_{\mu}}
+i(\hat{W}^-_{\mu}\frac{\delta\hat{\Gamma}}{\delta\hat{W}^-_{\mu}}
  -\hat{W}^+_{\mu}\frac{\delta\hat{\Gamma}}{\delta\hat{W}^+_{\mu}})
+i(\hat{\phi}^-\frac{\delta\hat{\Gamma}}{\delta\hat{\phi}^-}
  -\hat{\phi}^+\frac{\delta\hat{\Gamma}}{\delta\hat{\phi}^+})=0,
\end{array}
\end{equation}
and the Hermitian conjugate of Eqn. $(12)$. The above equations
are the starting point of all identities. For example, we have
in momentum space,
\begin{equation}
\begin{array}{l}
\displaystyle p^{\mu}\hat{\Gamma}^{\hat{A}\hat{W}^+\hat{W}^-}_{\mu\rho\sigma}
(p,p_+,p_-)
=e[\hat{\Gamma}^{\hat{W}^+\hat{W}^-}_{\rho\sigma}(-p_-)
  -\hat{\Gamma}^{\hat{W}^+\hat{W}^-}_{\rho\sigma}(p_+)],
\end{array}
\end{equation}
\begin{equation}
\begin{array}{l}
\displaystyle p^{\rho}_{+}\hat{\Gamma}^{\hat{A}\hat{W}^+\hat{W}^-}
                       _{\mu\rho\sigma}(p,p_+,p_-)
-M_W (1-\frac{\delta v}{v})
     \hat{\Gamma}^{\hat{A}\hat{\phi}^+\hat{W}^-}_{\mu\sigma}
                         (p,p_+,p_-) \\
\displaystyle =e[\hat{\Gamma}^{\hat{A}}_{\mu\sigma}(p)
  +\frac{c}{s}\hat{\Gamma}^{\hat{A}\hat{Z}}_{\mu\sigma}(p)
  -\hat{\Gamma}^{\hat{W}^+\hat{W}^-}_{\mu\sigma}(-p_-)].
\end{array}
\end{equation}
At tree level the above are easily checked. At one loop level they are
separately satisfied by counterterms and unrenormalized one loop
contributions. For the former the presence of $\delta v/v$
and Eqn. $(8)$ are crucial. Typical examples for the latter are shown
in Figs. 1 and 2.

We have discussed the feasibility of the BFM in spontaneously broken
gauge theories. In comparison to unbroken gauge theories new
features appear that are peculiar to spontaneously
broken gauge theories. These are associated with the presence of
unphysical Goldstone bosons and the  renormalizations
of masses generated by spontaneous symmetry breaking, making
the applicability of the BFM less evident.
We have shown by explicit one loop calculations
that the method can be consistently carried through in the on-shell
renormalization scheme of the electroweak theory.  Actually
the method is consistent with any scheme which treats the QED
subpart normally, i.e. ,with the electromagnetic coupling defined
in the Thomson limit and the $U(1)_{\rm e.m.}$ symmetry explicitly preserved
all the way. ( By the way we may mention that the
$\hat{A}_{\mu}-\hat{\phi}_{1,2}$ mixing is identically zero in the BFM. )
Due to the masslessness of unphysical Goldstone bosons
( we have actually verified this in general quantum $\xi$ gauges )
the choice of the background gauge ( parameterized by $\hat{\xi}_i$
and independent of the quantum gauge parameters $\xi_i$ ) in S matrix
calculations cannot be arbitrary and must be fixed
in the Landau gauge so that external propagators may be appropriately
amputated, if the S matrix involves external unphysical
Goldstone bosons. ( Indeed this is not the S matrix in the exact
sense though the usage prevails in the literature. )

The advantage of the BFM over the conventional approach is clear.
Since the Slavnov-Taylor identities are replaced by the naive Ward
identities, the renormalization structure is simplified and less
independent renormalization constants are needed. Technically,
calculations in the 't Hooft-Feynman gauge of the BFM are easier
than in the conventional approach. It is reasonable to expect that
this simplicity can be best enjoyed in even higher order calculations.
Since the relations between $1$PI functions are much simplified
in the BFM, we expect that some processes, which involve formidably
complicated gauge cancellations in the conventional approach,
e.g. , the longitudinal gauge boson scattering in the electroweak
theory, can be more easily computed in the BFM. The BFM may also
find its applications in the electroweak chiral Lagrangians
where gauge non-invariant terms are generally involved when
calculations are carried out to higher orders in general $R_{\xi}$
gauges of the conventional method. Work on these aspects is now
in progress.

We thank Prof. Y. P. Kuang for suggestions.

\begin{flushleft}
{\Large\sf Appendix}
\end{flushleft}

Some notations are listed in this appendix.
\begin{equation}
\begin{array}{l}
\displaystyle
Z_{\mu}=c~W^3_{\mu}-s~B_{\mu},~~~A_{\mu}=s~W^3_{\mu}+c~B_{\mu},\\
\displaystyle
\hat{Z}_{\mu}=c~\hat{W}^3_{\mu}-s~\hat{B}_{\mu},
{}~~\hat{A}_{\mu}=s~\hat{W}^3_{\mu}+c~\hat{B}_{\mu},
\end{array}
\end{equation}
where $c$ and $s$ are defined by renormalized couplings,
\begin{equation}
\displaystyle
c=\frac{g_2}{\sqrt{g^2_1+g^2_2}},~~~s=\frac{g_1}{\sqrt{g^2_1+g^2_2}}.
\end{equation}
Some renormalization constants appearing in the text are ( $i=1,2$ )
\begin{equation}
\begin{array}{l}
\displaystyle
Z^Z_i=c^2~Z^W_i+s^2~Z^B_i,~~Z^A_i=s^2~Z^W_i+c^2~Z^B_i,~~\\
\displaystyle
\delta Z^{AZ}_i=cs(\delta Z^{W}_i-\delta Z^{B}_i),\\
\displaystyle
\delta M^2_W=M^2_W(-2\frac{\delta v}{v}
                   +2\delta Z^W_1-3\delta Z^W_2+\delta Z_{\phi}),\\
\displaystyle
\delta M^2_Z=M^2_Z(-2\frac{\delta v}{v}
                   +2\delta Z^Z_1-3\delta Z^Z_2+\delta Z_{\phi}),\\
\displaystyle
\delta M^2_H=M^2_H(-3\frac{\delta v}{v}
                   +\frac{3}{2}\delta Z_{\lambda}-\delta Z_{\phi}
                   +\frac{\delta \mu^2}{M^2_H}),\\
\end{array}
\end{equation}


\newpage
\centerline{\large\bf Figure Captions }

\begin{enumerate}
\item A typical example of unrenormalized one loop contributions
      to Eqn. $(15)$. The virtual particles in loops are the quantum
      fields $W^{\pm},~Z$ or $A$. The solid ( dashed ) external lines
      are the background gauge ( unphysical Goldstone ) bosons.
\item A typical example of unrenormalized one loop contributions
      to Eqn. $(16)$.
\end{enumerate}

\end{document}